\documentclass[aps,prl,twocolumn,superscriptaddress]{revtex4}
\usepackage{graphicx}
\usepackage{color}

\begin{document}

\title{Quasi one-dimensional metallic band dispersion in the commensurate charge density wave of $1T$-TaS$_2$}
\author{Arlette S. Ngankeu}
\author{Kevin Guilloy}
\author{Sanjoy K. Mahatha}
\author{Marco Bianchi}
\author{Charlotte E. Sanders}
\affiliation{Department of Physics and Astronomy, Interdisciplinary Nanoscience Center (iNANO), Aarhus University, 8000 Aarhus C, Denmark}
\author{Kai Rossnagel}
\affiliation{Institute for Experimental and Applied Physics, University of Kiel, Germany}
\author{Jill A. Miwa}
\author{Philip Hofmann}
\email{philip@phys.au.dk}
\affiliation{Department of Physics and Astronomy, Interdisciplinary Nanoscience Center (iNANO), Aarhus University, 8000 Aarhus C, Denmark}
\date{\today}
\begin{abstract}
The commensurate charge density wave (CDW) in the layered compound $1T$-TaS$_2$ has hitherto mostly been treated as a quasi two-dimensional phenomenon. Recent band structure calculations have, however, predicted that the CDW coexists with a nearly one-dimensional metallic dispersion perpendicular to the crystal planes. Using synchrotron radiation based angle-resolved photoemission spectroscopy, we show that this metallic band does in fact exist. Its occupied band width is in excellent agreement with predictions for a simple $\tau_c$ stacking order of the CDW between adjacent layers and its periodicity in the $c$ direction is $2 \pi / c$.
\end{abstract}
\maketitle

The $1T$ polytype of  TaS$_2$ is one of the most studied layered transition metal dichalcogenides (TMDCs). Its rich electronic phase diagram involves several charge density wave (CDW) transitions driven by strong electronic correlations and electron-phonon coupling \cite{Wilson:1975aa,Rossnagel:2011aa}. Particular focus has been on the ground state below 180~K which is a commensurate CDW phase with a $\sqrt{13} \times \sqrt{13}$ so-called Star of David reconstruction that is rotated by $13.9^{\circ}$ against the lattice. In this phase, the large-scale periodic lattice distortion is thought to coexist with a Mott insulating ground state arising from the single electron localized on the centre atom of the Star of David \cite{Wilson:1975aa,Fazekas:1979aa}. While the research on CDWs in layered TMDCs is more than 40 years old, renewed  interest has been driven by the possibility to elucidate transitions between different CDW states using ultrafast techniques \cite{Perfetti:2006aa,Perfetti:2008aa,Hellmann:2010aa,Eichberger:2010aa,Petersen:2011ab,Stojchevska:2014aa}; by the observation of metastable ``hidden states'' \cite{Stojchevska:2014aa}; and by the experimental accessibility of metallic TMDCs as single layers \cite{Ugeda:2016aa,Sanders:2016aa}.

The realization that CDWs could be different in single layer TMDCs than in analogous bulk materials has drawn attention to the fact that viewing the bulk materials' electronic properties as essentially two-dimensional might be an oversimplification. While reduced dimensionality has a significant impact on electronic instabilities, due to increased electronic correlations and electron-phonon coupling, interlayer coupling also appears to be essential for a full understanding of the electronic properties of these materials \cite{Bovet:2003aa,Rossnagel:2005ab,Freericks:2009ab,Darancet:2014aa,Ritschel:2015aa,Lazar:2015aa}. Specifically, several calculations predict a one-dimensional metallic band formation along the $\Gamma-A$ direction of the Brillouin zone in the ground state CDW of $1T$-TaS$_2$ (i.e., perpendicular to the planes). This is found in density functional theory calculations  \cite{Bovet:2003aa,Ritschel:2015aa,Lazar:2015aa}, even when electronic correlations are taken into account \cite{Darancet:2014aa,Freericks:2009ab}. While the metallic band along  $\Gamma-A$ is universally found in all calculations, the details of the dispersion depend on the stacking order of the CDW unit cell between adjacent planes \cite{Darancet:2014aa,Ritschel:2015aa}. 

Angle-resolved photoemission spectroscopy (ARPES) is an experimental technique capable of determining the three-dimensional band structure of crystalline solids, and numerous ARPES studies have been performed on $1T$-TaS$_2$ (for a review see Ref. \cite{Rossnagel:2011aa}). However, very little attention has been paid to the possibly three-dimensional character of the band structure. In this Letter, we report a systematic determination of the band structure perpendicular to the planes of $1T$-TaS$_2$ from high-quality crystals in the commensurate CDW phase, with special focus on the possible metallicity of the compound in this direction. We do observe the theoretically predicted one-dimensional metallic band. The occupied band width and the observed periodicity agree with a simple $\tau_c$ stacking order of the CDW---i.e., a stacking in which the center atoms of the Stars of David are placed directly on top of each other in adjacent layers. 

The $1T$-TaS$_2$ crystals were grown from high purity elements by chemical vapor transport using iodine as a transport agent; for details see Ref. \cite{Haupt:2016aa}. The crystals were cleaved at $\approx$30~K in ultra-high vacuum, and ARPES data were collected on the SGM-3 end station of ASTRID2 \cite{Hoffmann:2004aa}.  The energy resolution varied from $\approx$50 to $\approx$130~meV for the lowest and highest photon energies, respectively. The angular resolution was better than 0.2$^{\circ}$. The temperature during the ARPES experiments was $\approx$30~K. 

\begin{figure}
\includegraphics[width=0.45\textwidth]{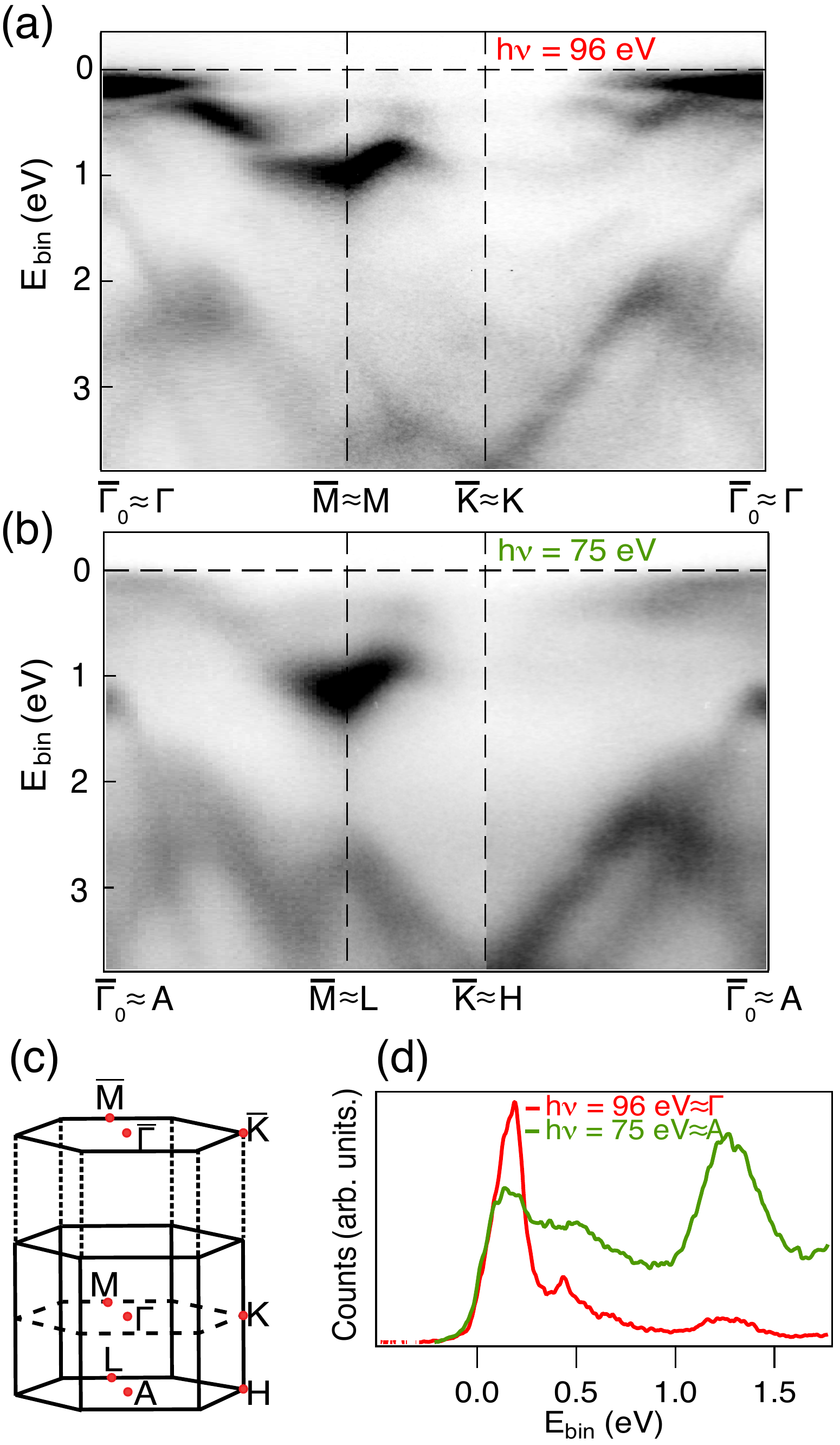}\\
\caption{(Color online) (a),(b) Photoemission intensity along high symmetry lines of the surface Brillouin zone for photon energies of 96 and 75~eV, respectively. Dark corresponds to high photoemission intensity. The high symmetry points in the surface Brillouin zone (noted with a bar over the letter) are reached exactly but the bulk high symmetry points only approximately (see text and Ref. \cite{SMAT}). (c)  Sketch of the first Brillouin zone of $1T$-TaS$_2$ and its projection on the (0001) surface. (d) Energy dispersion curves at the $\bar{\Gamma}$ point for photon energies of 96 (red) and 75~eV (green), corresponding to the bulk $\Gamma$ and $A$ points, respectively.}
  \label{fig:1}
\end{figure}

A first suggestion of the three-dimensional character of the $1T$-TaS$_2$ band structure in the commensurate CDW is given in Fig. \ref{fig:1}, which shows the photoemission intensity along high-symmetry directions in the surface Brillouin zone for two different photon energies (96 and 75~eV for Fig. \ref{fig:1}(a) and (b), respectively). Assuming free-electron final states and using a procedure outlined below \cite{SMAT}, the photon energies have been chosen such that data are collected approximately in the $\Gamma-M-K$ plane and $A-L-H$ plane of the bulk Brillouin zone for (a) and (b), respectively. For a definition of the high-symmetry points see Fig. \ref{fig:1}(c). Note, however, that only the parallel component of the crystal momentum $k_{\parallel}$ is conserved in the photoemission process. The high-symmetry points of the \emph{surface} Brillouin zone are reached exactly.  The $\Gamma$ and $A$ points in the \emph{bulk} Brillouin are also reached rather precisely, but the $M,K,L,H$ points at finite $k_{\parallel}$  only approximately \cite{SMAT}.

At first glance, the dispersions are very similar to each other and in good agreement with previous ARPES results \cite{Pillo:1999ac,Arita:2004aa,Bovet:2004ab,Perfetti:2005aa,Rossnagel:2005ab,Clerc:2006aa,Ritschel:2015aa}. The states close to the Fermi energy $E_F$ are broad due to the strongly correlated character of the material,  with a lack of any clear Fermi level crossings. However, upon closer inspection, some differences between Fig. \ref{fig:1}(a) and (b) can be noted. The deeper lying, sulphur 2p-derived \cite{Mattheiss:1973aa,Manzke:1988aa} states are expected to be less two-dimensional and do indeed show a different dispersion, for example around $\bar{M}$. The states near $E_F$, on the other hand, mostly differ in their intensity. Note, for example that the flat band immediately below $E_F$ near $\Gamma$ in Fig. \ref{fig:1}(a) is very well defined, while it is almost absent at $A$. This difference is best seen in a direct comparison of energy distribution curves through $\Gamma$ and $A$, as given in Fig. \ref{fig:1}(d). 

\begin{figure}
\includegraphics[width=0.45\textwidth]{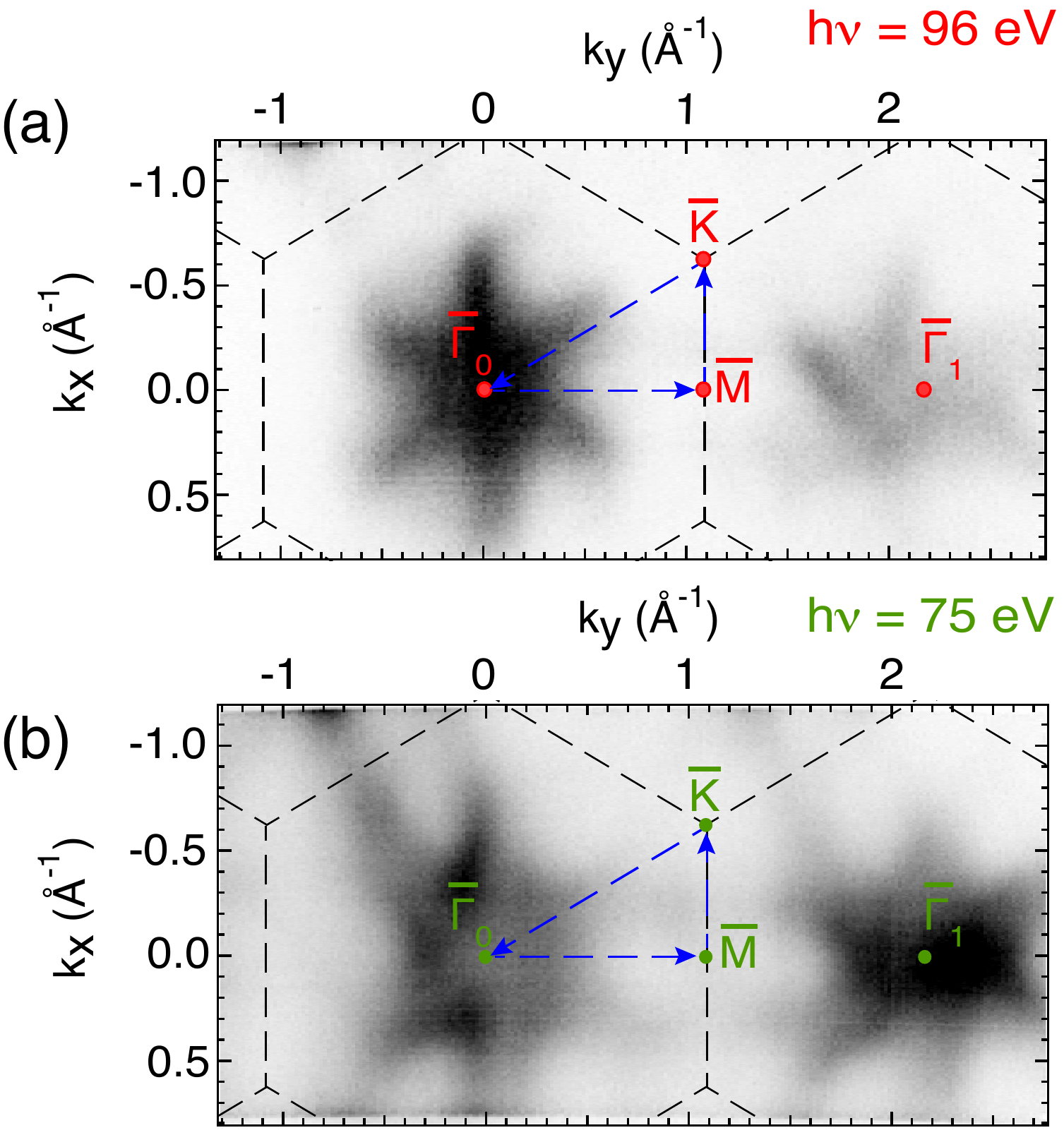}\\
\caption{(Color online) (a),(b) Photoemission intensity 90~meV below the Fermi energy for photon energies of  96 and 75~eV, respectively. $\bar{\Gamma}_0$ corresponds to normal emission and to the bulk $\Gamma$ and $A$ points in (a) and (b) respectively. }
  \label{fig:1b}
\end{figure}

Subtle differences in the states near the Fermi level are especially well seen in the $k_{\parallel}$-dependence of the photoemission intensity. Fig. \ref{fig:1b}(a) and (b) show such intensity plots at a binding energy of 90~meV, taken for the same photon energies as  the data in Fig. \ref{fig:1}(a) and (b). The plots show distinct differences. In particular, the photoemission intensity reaches a maximum at normal emission (marked as $\bar{\Gamma}_0$) in (a) while it shows a local minimum in (b), indicating a dependence of the electronic structure on the perpendicular crystal momentum  $k_{\perp}$. 

\begin{figure}
\includegraphics[width=0.5\textwidth]{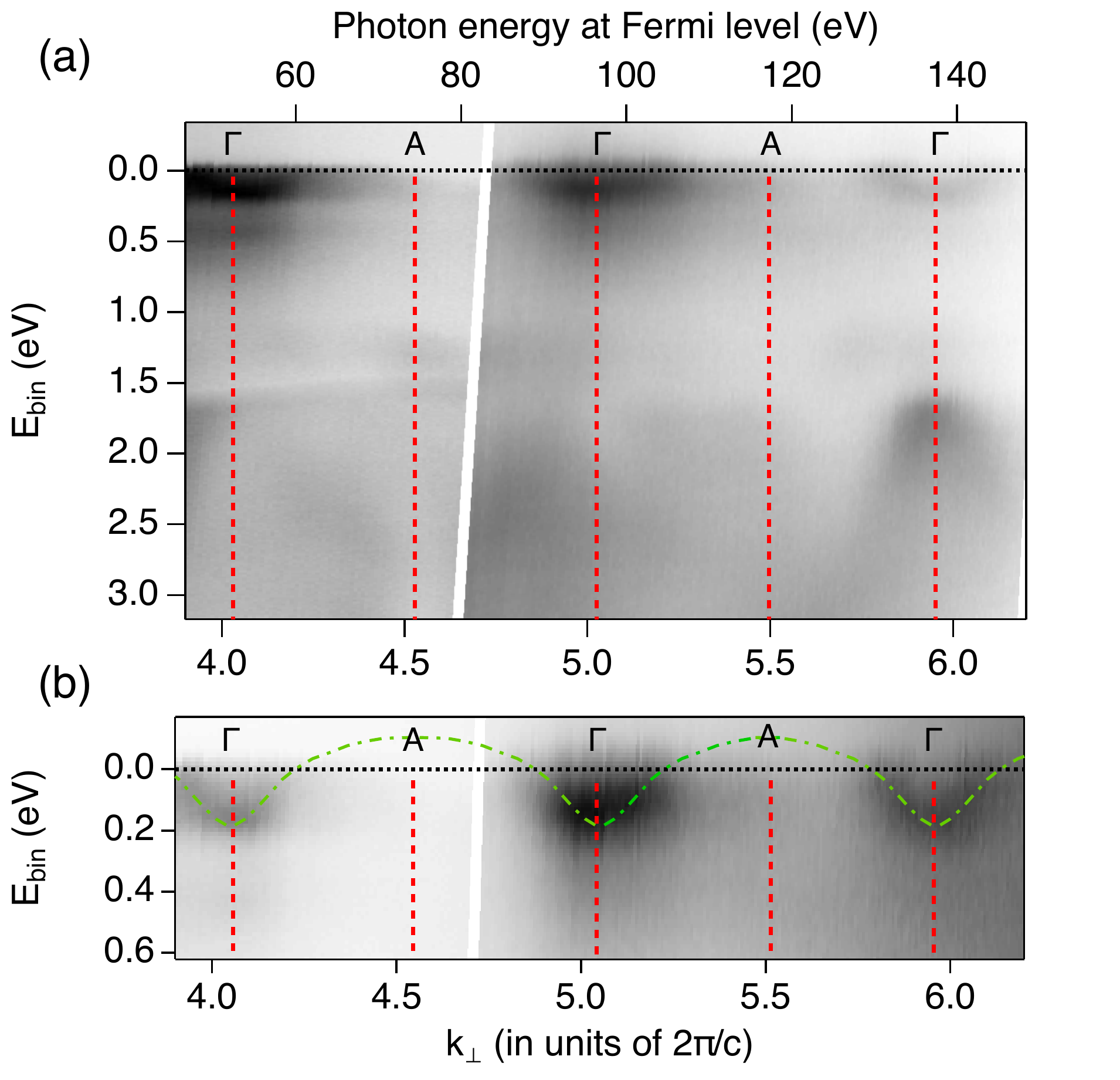}\\
\caption{(Color online) (a) Photoemission intensity measured in normal emission as a function of photon energy $h\nu$, here converted to $k_{\perp}$ using free electron final states. $k_{\perp}$ values are given in units of the reciprocal lattice vector $2 \pi / c$. The $h\nu$ values given on the upper axis refer to the photon energy for the states at the Fermi energy. The greyscale is logarithmic. The dashed red lines mark the maximum binding energy of the small electron pocket near $\Gamma$ and the $k_{\perp}$ values mid-way between two $\Gamma$ points. (b) Magnification of the intensity in the vicinity of the Fermi energy. The intensity is normalized by an exponential function of the photon energy. The dashed-dotted green lines are the result of the calculation from Ref. \cite{Darancet:2014aa}.}
  \label{fig:2}
\end{figure}

\begin{figure}
\includegraphics[width=0.5\textwidth]{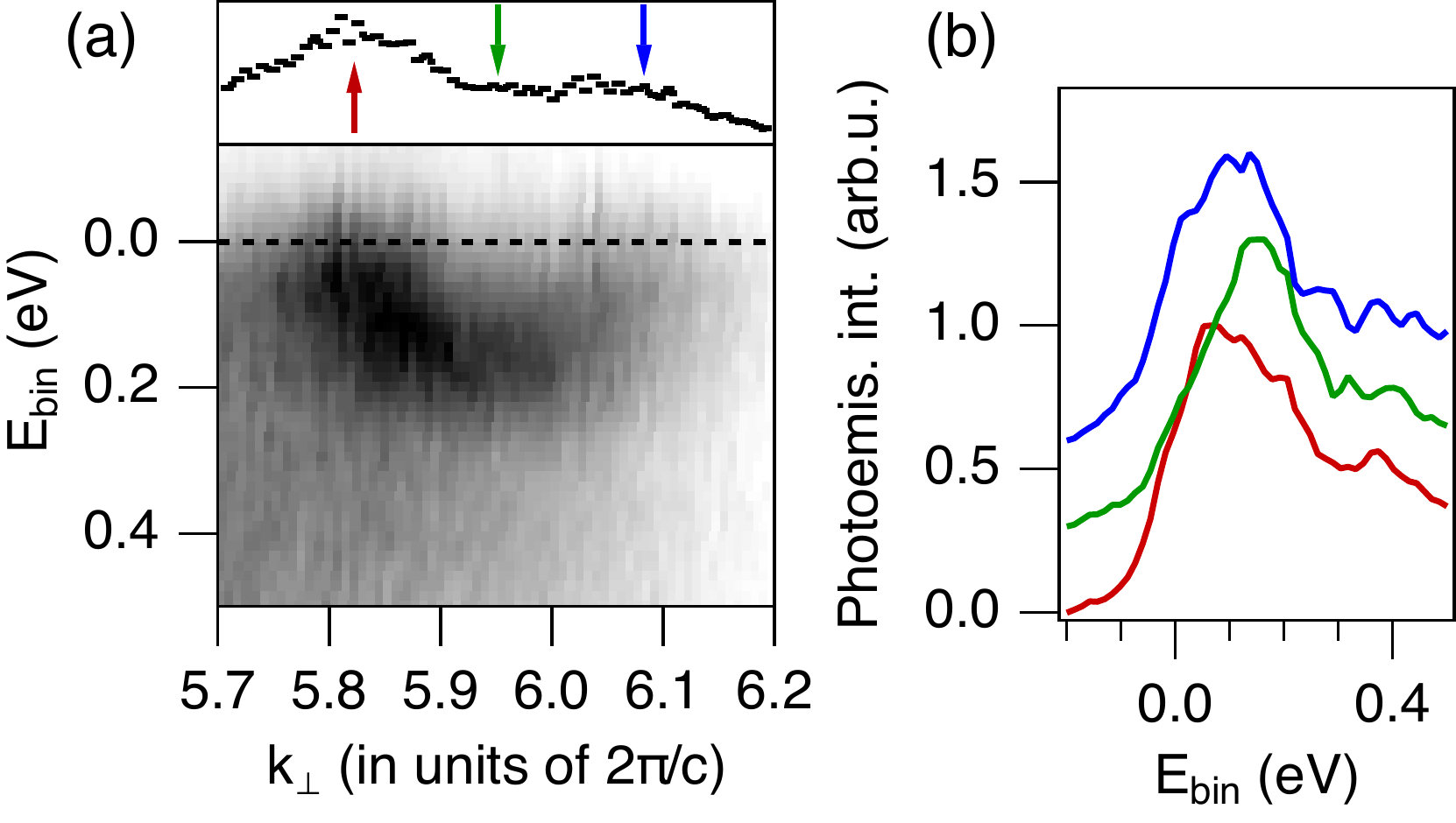}\\
\caption{(Color online) (a) Magnification of the data in Fig. \ref{fig:2} around $k_{\perp}=12\pi /c$ with a momentum distribution curve ($k$-dependent photoemission intensity) at the Fermi energy on the top. (b) Energy distribution curves taken at the arrows in (a), i.e. near the Fermi level crossings and at maximum binding energy of the band. The curves are normalized and vertically displaced. }
  \label{fig:4}
\end{figure}

The possible existence of the predicted quasi one-dimensional band along $\Gamma-A$ can be established by collecting the photoemission intensity in normal emission as a function of photon energy $h\nu$ and binding energy $E_{bin}$. The result of such a scan is shown in Fig. \ref{fig:2}(a). The photoemission intensity $I$ in the figure has been converted from the raw data ($I$ measured as a function of $h\nu$ and $E_{bin}$) to a function of  $k_{\perp}$  and $E_{bin}$, using the assumption of free-electron final states \cite{SMAT}. The most important feature in the data is the small electron pocket appearing near the Fermi energy around the $\Gamma$ points, i.e. at integer multiples of $k_{\perp}= 2 \pi / c$ ($c=5.86$~{\AA} \cite{Wilson:1969}). For clarity, the intensity close to $E_F$ is magnified  in Fig. \ref{fig:2}(b) and a  detailed view of the situation at the highest $k_{\perp}$ is given in Fig. \ref{fig:4}. The experimental observation of this metallic band is the central result of this paper.

It should be noted that recovering an approximate initial state $k_{\perp}$ using free-electron final states requires a choice of the solid's inner potential $V_{0}$ and work function $\Phi$ \cite{Plummer:1982aa}. Here $V_{0}=20$~eV and $\Phi=4.5$~eV were chosen in order to place the periodically appearing electron pocket close to the $\Gamma$ point of the Brillouin zone. While somewhat different values for $V_0$ have been used in very early investigations \cite{Manzke:1988aa}, we emphasize that the precise choice of this parameter is not critical. Indeed, due to the symmetry of the electron pocket's dispersion, it can only be placed at either $\Gamma$ or $A$ and no choice of $V_0$ below 40~eV would lead to the electron pocket being found at $A$. Moreover, the choice of the inner potential is also confirmed by the symmetry of the deeper lying bands, especially the sulphur p-bands that show a binding energy minimum at $\Gamma$ \cite{Mattheiss:1973aa}. This minimum is most clearly identified at a binding energy of $\approx$1.7~eV for $k_{\perp}= 12 \pi / c$. 

Fig. \ref{fig:2}(b) also shows the calculated band structure from Ref. \cite{Darancet:2014aa} superimposed on the data as dashed lines. The agreement is excellent. Other published calculations show very similar dispersions \cite{Bovet:2003aa,Freericks:2009ab,Ritschel:2015aa,Lazar:2015aa}. Moreover, the observed $k_{\perp}$ periodicity of $2 \pi / c$ suggests a real space periodicity of only one unit cell and hence the $\tau_c$ stacking. This is consistent with  calculations, most of which were made under the assumption of  $\tau_c$ stacking.

On the other hand, a regular $\tau_c$ stacking does not agree with a substantial amount of structural information available on the commensurate CDW phase---see, e.g., Refs. \cite{Tanda:1984aa,Nakanishi:1984aa,Ishiguro:1991aa}. Data from different techniques reveal considerable disorder in the $c$ direction, accompanied by different stackings. This does not, however,  imply a contradiction between the ARPES observations and the structural data. First of all, only periodic contributions to the stacking give rise to any regular band structure, while disorder merely increases the background intensity. Moreover, we note that the observed bands are rather broad, as seen in the momentum and energy distribution curves of Fig. \ref{fig:4}, suggesting that $k_{\perp}$ is not well defined. This can be partly due to disorder, in addition to the intrinsic uncertainty in $k_{\perp}$ that stems from the short inelastic mean free path of electrons in solids and the localisation accompanying this \cite{Plummer:1982aa}. 

An alternative way to view the $2 \pi / c$ periodicity is that,  in the absence of a CDW, this would be expected for every band in the $1T$ polytype. Since the CDW is, after all, only a minor distortion of the lattice \cite{Rossnagel:2011aa}, such a periodicity could still be present. This interpretation is supported by an accurate band structure calculation for the undistorted $1T$ structure (including the significant spin-orbit coupling \cite{Rossnagel:2006aa}) that shows a very similar metallic dispersion in the $\Gamma-A$ direction, even though the in-plane dispersion is completely different from the CDW case \cite{Darancet:2014aa}. 

The observed electron pocket around $\Gamma$ also appears to explain the distinct differences in the constant energy surfaces of Fig. \ref{fig:1b}, since the band is occupied at  $\Gamma$ but empty at $A$. Indeed, Ritschel \emph{et al.} have challenged the common view that this band is the lower Hubbard band of the Mott insulating state because it can  be reproduced by a calculation not including correlations \cite{Ritschel:2015aa}. However, the situation is more complex because the metallic band dispersion appears to coexist with a part of the spectral weight  at the original peak position of the lower Hubbard band, as seen in the energy distribution curves of Fig. \ref{fig:1}(d). Indeed,  the peak that is usually assigned to the lower Hubbard band significantly changes its intensity but never entirely disappears, not even at the $A$ point of the bulk Brillouin zone (corresponding to h$\nu$=75~eV) where the strongly dispersing metallic band is predicted to be well above the Fermi level. Such a complex behaviour is not unexpected given the partially disordered character of the CDW along $c$ \cite{Tanda:1984aa,Nakanishi:1984aa,Ishiguro:1991aa}, which should limit the formation of a well-defined band structure in this direction. 

The observation of a metallic band is partly consistent with the reported transport phenomena in the material. In the temperature region immediately below the transition to the commensurate CDW ($\approx$50 -- 80~K), a metallic temperature dependence of the resistivity is observed, consistent with the remaining metallic band, but at very low temperature the resistivity increases strongly \cite{Salvo:1977aa}. This has been ascribed to disorder-induced Anderson localization \cite{Salvo:1977aa}, an interpretation that appears consistent with the observed disorder in the $c$ direction \cite{Tanda:1984aa,Nakanishi:1984aa,Ishiguro:1991aa}.

Given the one-dimensional metallic dispersion, one would expect that the resistivity $\rho_c$ in the $c$ direction would be lower than the resistivity $\rho_a$ in the plane, but the opposite is found. In fact, a direct measurement of $\rho_c / \rho_a$ gives a value of approximately 500 \cite{Hambourger:1980aa},  even in the temperature range of metallic conductance. The apparent contradiction of a metallic band with a lack of metallic conduction could be due to a gap opening in the one-dimensional dispersion near $E_F$. This is not supported by the detailed view on the dispersion in Fig. \ref{fig:4} which appears to show clear Fermi level crossings. However, the rather broad features do not allow us to draw a definite conclusion about this type of gap formation. Moreover, the simultaneous presence of the dispersing band and the lower Hubbard band throughout the bulk Brillouin zone increase the difficulty of identifying a clear gap opening.

In conclusion, we have observed a one-dimensional metallic band in the $c$ direction of $1T$-TaS$_2$. This has recently been predicted by several calculations but hitherto never been observed. The result gives strong experimental support to the notion that viewing the TMDC CDW materials as quasi two-dimensional is an oversimplification. It also suggests that new rich physics can be expected from truly two-dimensional single layers of these compounds, not only because of the absence of interaction with neighboring crystal planes but also because the electronic properties can be influenced by substituting these planes with other materials of choice.

This work was supported by the Danish Council for Independent Research, Natural Sciences under the Sapere Aude program (Grants No. DFF-4002-00029 and DFF-6108-00409) and by VILLUM FONDEN via the Centre of Excellence for Dirac Materials (Grant No. 11744).

\end{document}